\documentclass{article}

\usepackage{graphicx}
\usepackage{amsmath,graphicx,amssymb,amsfonts}
\usepackage{color}
\usepackage[normalem]{ulem}

\title{\bf Map Projection}
\author{Ebrahim Ghaderpour \\[0.3cm]
\small{ Email: ebig2@yorku.ca} \\
\small{ Department of Earth and Space Science and Engineering,}\\
\small{York University, Toronto, Canada}\\[0.2cm]}
\date{}
\newtheorem{preex}{{\bf Example}}

\newenvironment{example}[1]{\begin{preex}{\hspace{-0.05
               em}{\rm #1}}}{\end{preex}}

\begin{document}
\maketitle

\begin{abstract}
\noindent In this paper, we introduce some known map projections from a model of the Earth to a flat sheet of paper or map and derive the plotting equations for these projections. The first fundamental form and the Gaussian fundamental quantities are defined and applied to obtain the plotting equations and  distortions in length, shape and size for some of these map projections.   
\end{abstract}

The concepts, definitions and proofs in this work are chosen mostly from \cite{P, RA}. 

\section{Introduction}

 A {\it map projection} is a systematic transformation of the latitudes and longitudes of positions on the surface of the Earth to a flat sheet of paper, a map. More precisely, a map projection requires a transformation from a set of two independent coordinates on the model of the Earth (the latitude $\phi$ and  longitude $\lambda$)  to a set of two independent coordinates on the map (the Cartesian coordinates $x$ and $y$), i.e., a transformation matrix $T$ such that $$\left[ \begin {array}{c} x\\ y\end {array} \right]=T\left[ \begin {array}{c} \phi \\ \lambda\end {array} \right]\!.$$  However, since we are dealing with partial derivative and fundamental quantities (to be defined later), it is impossible to find such a transformation explicitly. 

There are a number of techniques for map projection, yet in all of them distortion occurs in length, angle, shape, area or in a combination of these. Carl Friedrich Gauss showed  that a sphere's surface cannot be represented on a map without distortion (see \cite{P}). 

A {\it terrestrial globe} is a three dimensional scale model of the Earth that does not distort the real shape and the real size of  large futures of the Earth. The term {\it globe} is used for those objects that are approximately spherical. 
The equation for spherical model of the Earth with radius $R$ is 
\begin{align}\label{spherical}
\dfrac{x^2}{R^2}+\dfrac{y^2}{R^2}+\dfrac{z^2}{R^2}=1.
\end{align}

An  oblate {\it ellipsoid} or {\it spheroid} is a quadratic surface obtained by rotating an ellipse about its minor axis (the axis that passes through the north pole and the south pole).
The shape of the Earth is appeared to be an oblate ellipsoid (mean Earth ellipsoid), and the geodetic latitudes and longitudes of positions on the surface of the Earth coming from satellite observations are on this ellipsoid.
The equation for spheroidal model of the Earth is 
\begin{align}\label{spheroidal}
\dfrac{x^2}{a^2}+\dfrac{y^2}{a^2}+\dfrac{z^2}{b^2}=1,
\end{align}
 where $a$ is the semimajor axis, and $b$ is the semiminor axis of the spheroid of revolution. 

The spherical representation of the Earth (terrestrial globe) must be modified to maintain accurate representation of either shape or size of the spheroidal representation of the Earth. We discuss about these two representations in Section \ref{section4}.

There are three major types of map projections:

{\bf 1. Equal-area projections.} These projections preserve the area (the size) between the map and the model of the Earth. In other words, every section of the map keeps a constant ratio to the area of the Earth which it represents. Some of these projections are Albers with one or two standard parallels (the conical equal-area), the Bonne, the azimuthal and Lambert cylindrical equal-area which are best applied 
to a local area of the Earth, and some of them are world maps such as the sinusoidal, the Mollweide, the parabolic, the Hammer-Aitoff, the Boggs eumorphic, and Eckert IV.

{\bf 2. Conformal projections.}  These projections maintain the shape of an area during transformation from the Earth to a map. These projections include the Mercator, the Lambert conformal with one standard parallel, and the stereographic. These projections are only applicable to limited areas on the model of the Earth for any one map. Since there is no practical use for conformal world maps, conformal world maps are not considered.

{\bf 3. Conventional projections.} These projections are neither equal-area nor conformal, and they are designed based on some particular applications. Some examples are the simple conic, the gnomonic, the azimuthal equidistant, the Miller, the polyconic, the Robinson, and the plate carree projections.

In this paper, we only show the derivation of plotting equations on a map for the Mercator and Lambert cylindrical equal-area for a spherical model of the Earth (Section \ref{section2}), 
the Albers  with one standard parallel  and the azimuthal for a spherical model of the Earth and  the Lambert conformal  with one standard parallel for a spheroidal model of the Earth (Section \ref{section5}), the sinusoidal (Section \ref{section6}), the simple conic and the plate carree projections (Section \ref{section7}). The methods to obtain other projections are similar to these projections, and the reader is referred to \cite{DA, P, RA}.

Suppose that a terrestrial glob is covered with infinitesimal circles. In order to show distortions in a map projection,  one may look at the projection of these circles in a map which are ellipses whose axes are the two principal directions along which scale is maximal and minimal at that point on the map. This mathematical contrivance is called {\it Tissot's indicatrix}.

Usually Tissot's indicatrices are placed across a map along the intersections of meridians and parallels to the equator, and they provide a good tool to calculate the magnitude of distortions at those points (the intersections). 

In an equal-area projection, Tissot's indicatrices change shape (from circles to ellipses), whereas their areas remain the same. In conformal projection, however, the shape of circles preserves, and the area varies. In conventional projection, both shape and size of these circles change. In this paper, we portrayed the Mercator, the Lambert cylindrical equal-area,  the sinusoidal and the plate carree maps with Tissot's indicatrices.

In Section \ref{section8}, the equations for  distortions of length, area and angle are derived, and distortion in length for the Albers projection and in length and area for the Mercator projection are calculated, \cite{P, RA, VK}.

\section{Mercator projection and Lambert cylindrical projection}\label{section2}
In this section, by an elementary method, we show the cylindrical method that Mercator used to map from a spherical model of the Earth to a flat sheet of paper. Also, we give the plotting equations for the Lambert cylindrical equal-area projection. Then, in Section \ref{section3}, we obtain  the Gaussian fundamental quantities, and show a routine mathematical way to find plotting equations for different map projections.  This section is based mostly on \cite{O}.

\begin{figure}[!htp] 
\centering
\includegraphics[height=8cm, width=11cm]{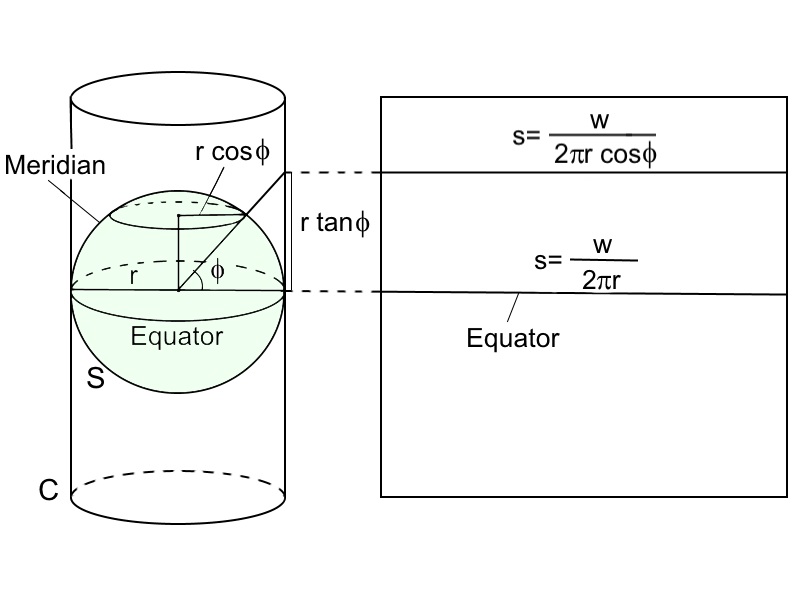}
\caption{  Geometry for the cylindrical projection}
\label{cylinder}
\end{figure}

Let $S$ be the globe, and $C$ be a circular cylinder tangent to  $S$ along the equator, see Fig. \ref{cylinder}. Projecting $S$ along the rays passing through the center of $S$ onto $C$, and unrolling the cylinder onto a vertical strip in a plane is called {\it central cylindrical projection}. Clearly, each meridian on the sphere is mapped to a vertical line to the equator, and each parallel of the equator is mapped onto a circle on the cylinder and so a line parallel to the equator on the map.
All methods discussed in this section and other sections are about central projection, i.e., rays pass through the center of the Earth to a cone or cylinder.  Methods for those projections that are not central are similar to central projections (see \cite{P, RA}).

Let $w$ be the width of the map. The scale of the map along the equator is  $s=w/(2\pi R)$ that is the ratio of size of objects drawn in the map to actual size of the object it represents. The scale of the map usually is shown by three methods: arithmetical (e.g. 1:6,000,000), verbal (e.g. 100 miles to the inch) or geometrical.

At latitude $\phi$, the parallel to the equator is a circle with circumference $2\pi R \cos \phi$, so the scale of the map at this latitude is 
\begin{align}\label{sh}
s_h=\dfrac{w}{2\pi R \cos\phi}=s\sec\phi, 
\end{align}
where the subscript $h$ stands for horizontal.

Assume that $\phi$ and $\lambda$ are in radians, and the origin in the Cartesian coordinate system corresponds to the intersection of the Greenwich meridian ($\lambda=0$) and 
the equator ($\phi=0$). Then every cylindrical projection is given explicitly by the following equations
\begin{align}\label{xymercator}
x=\dfrac{w\lambda}{2\pi}, \ \ \ \ \ y=f(\phi).
\end{align}
For instance, it can be seen from Fig. \ref{cylinder}  that  a central cylindrical  projection is given by 
$$x=\dfrac{w\lambda}{2\pi}, \ \ \ \ \ y=r\tan\phi,$$  
where for a map of width $w$, a globe of radius $r=w/(2\pi)$ is chosen.

In a globe, the arc length between latitudes of $\phi$ and $\phi_1$ (in radians) along a meridian is $$2\pi R\cdot \dfrac{\phi_1-\phi}{2\pi}=R(\phi_1-\phi),$$ and the image on the map has the length $f(\phi_1)-f(\phi)$. So the overall scale factor of this arc along the meridian when $\phi_1$ gets closer and closer to $\phi$ is 
\begin{align}\label{sv}
s_{v}=\dfrac{1}{ R}f'(\phi)=\dfrac{1}{ R}\lim_{\phi_1 \rightarrow \phi}\dfrac{f(\phi_1)-f(\phi)}{\phi_1-\phi},
\end{align}
where the subscript $v$ stands for vertical.

\begin{figure}[!htp]
\centering
\begin{minipage}{.5\textwidth}
  \centering
  \includegraphics[height=6.5cm, width=7.8cm]{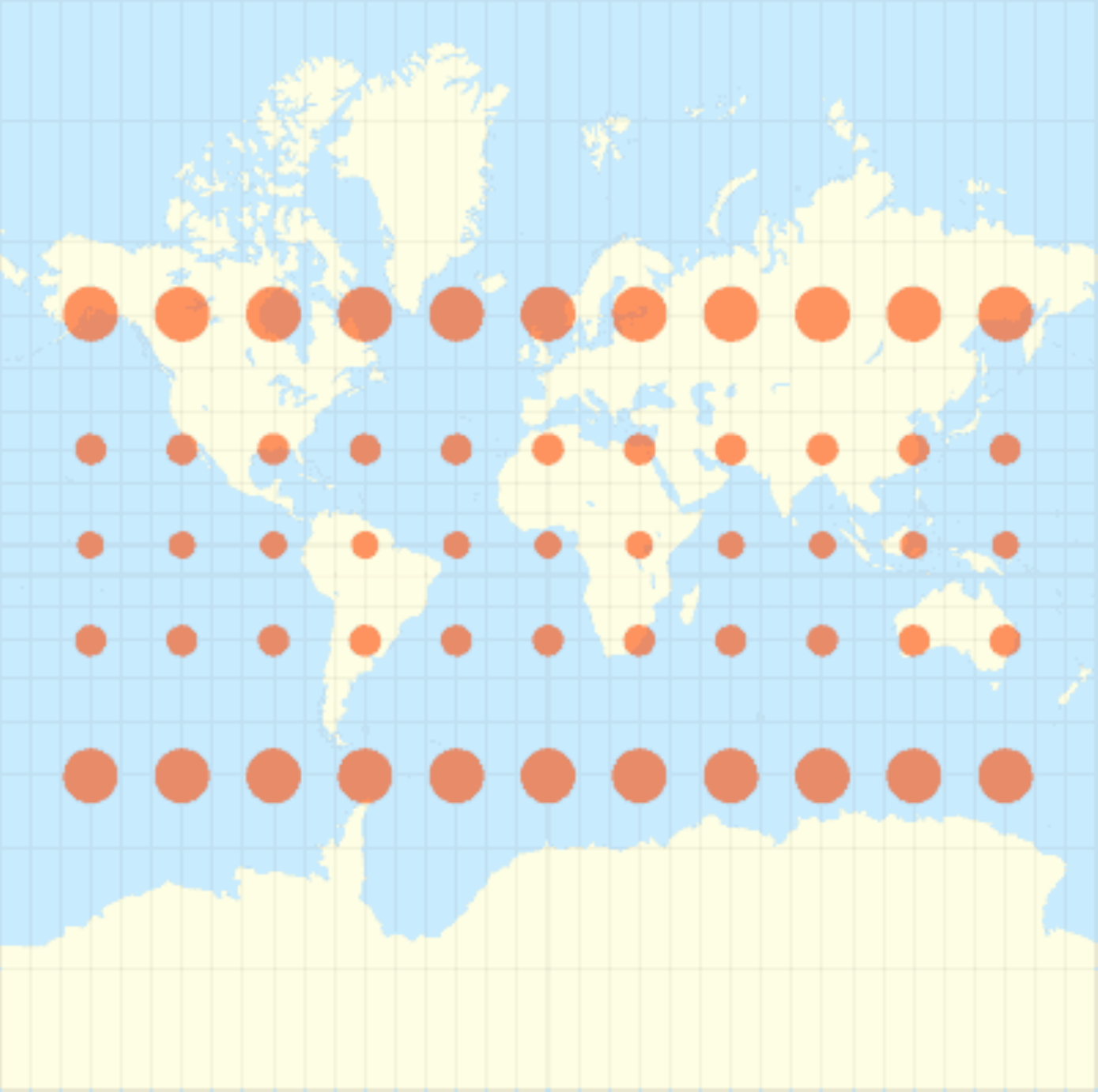}
  \caption{The Mercator conformal map}
  \label{mer11}
\end{minipage}%
\begin{minipage}{.5\textwidth}
  \centering
  \includegraphics[height=6.5cm, width=7.8cm]{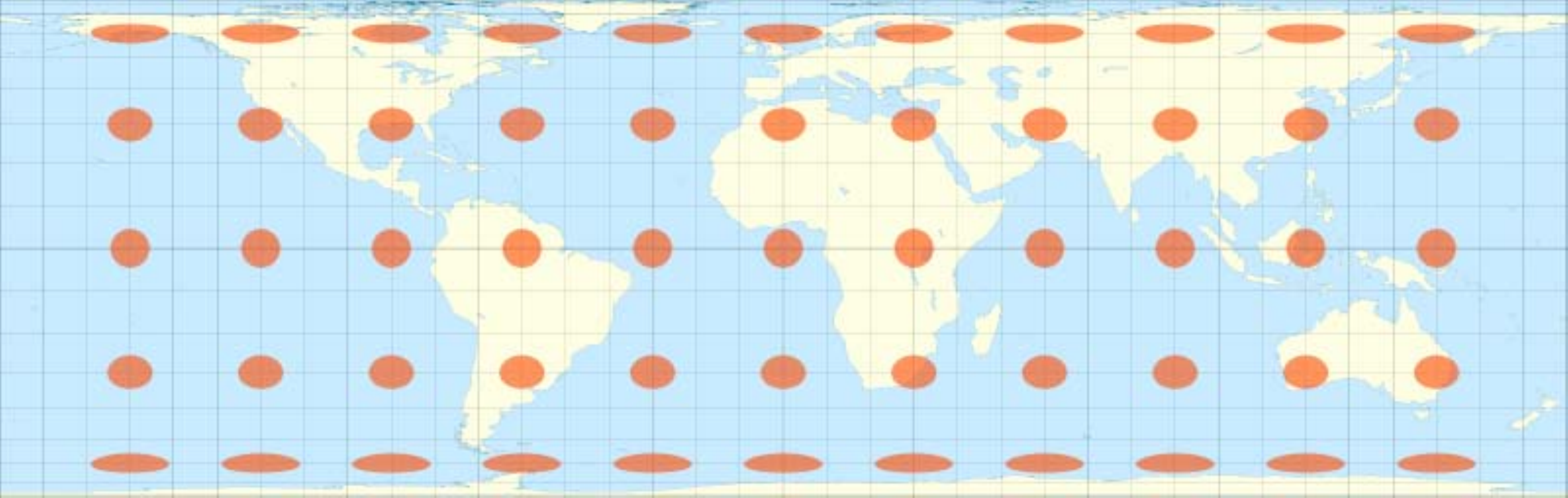}
  \caption{The Lambert equal-area map}
  \label{lam11}
\end{minipage}
\end{figure}

The goal of Mercator was to equate the horizontal scale with vertical scale at  latitude $\phi$, i.e., $s_h=s_v$. Thus, from Eqs. \eqref{sh} and \eqref{sv}, 
\begin{align}\label{conmer}
f'(\phi)=\dfrac{w}{2\pi}\sec \phi.
\end{align}
Mercator was not be able to solve Equation \ref{conmer} precisely because logarithms were not invented! 
But now, we know that the following is the solution to Eq. \eqref{conmer} (use $f(0)=0$ to make the constant coming out from the integration equal to zero), 
\begin{align}\label{ymer}
y=f(\phi)=\frac{w}{2\pi}\ln |\sec\phi +\tan\phi|.
\end{align}
Thus, the equations for the Mercator conformal projection (central cylindrical conformal mapping) are $$x=\dfrac{w\lambda}{2\pi}, \ \ \ \ \ y=\frac{w}{2\pi}\ln |\sec\phi +\tan\phi|.$$ 
Fig. \ref{mer11} shows the Mercator projection with Tissot's indicatrices that do not change their shape (all of them are circles indicating a conformal projection) while their size get larger and larger toward the poles.


Now if the goal is preserving size rather than shape, then we would make the horizontal and vertical scaling reciprocal, so the stretching in one direction will match shrinking in the other. Thus, from Eqs. \eqref{sh} and \eqref{sv}, we obtain $f'(\phi)\sec\phi=c$ or 
\begin{align}\label{eqmer}
f'(\phi)=c\cos\phi,
\end{align}
where $c$ is a constant. From Eqs. \eqref{conmer} and \eqref{eqmer}, we can choose $c$ in such away that for a given latitude, the map also preserves the shape in  that area. For instance if $\phi=0$, then we choose $c=w/(2\pi)$, and so the map near equator is conformal too.  
Hence, the equations for the cylindrical equal-area projection (one of Lambert's maps) are $$x=\dfrac{w\lambda}{2\pi}, \ \ \ \ \ y=\dfrac{w}{2\pi}\sin\phi.$$ 
Fig. \ref{lam11} shows the Lambert projection with Tissot's indicatrices that do not change their size (indicating an equal-area projection) while their shape are changing toward the poles.

\section{First fundamental form}\label{section3}
In this section, we derive the first fundamental form for a general surface that completely describes the metric properties of the surface, and it is a key in map projection, \cite{G, P, RA, VK}.

\begin{figure}[!htp] 
\centering
\includegraphics[height=6cm, width=9cm]{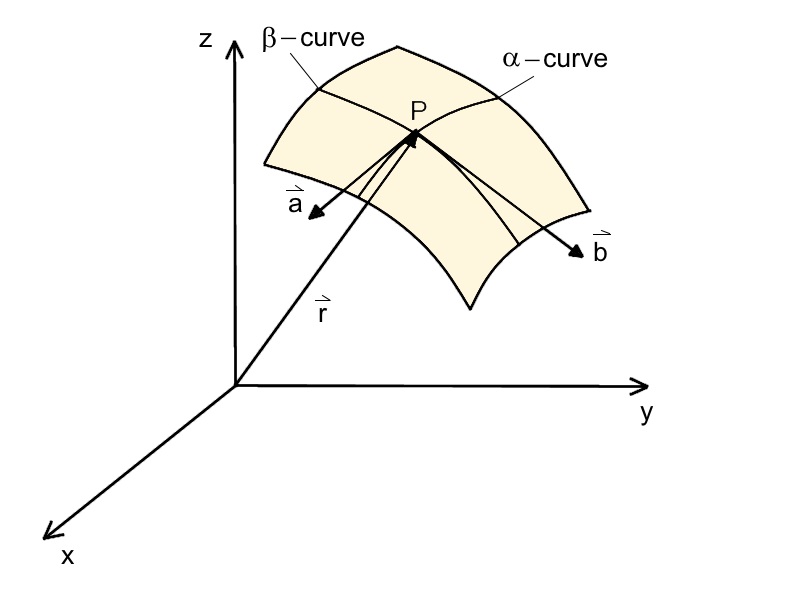}
\caption{  Geometry for parametric curves }
\label{ff}
\end{figure}

The vector at any point $P$ on the surface is given by $\overrightarrow{r}=\overrightarrow{r}(\alpha, \beta)$. If either of parameters $\alpha$ or $\beta$ is held constant and the other one is varied, a space curve results, see Fig. \ref{ff}.

The tangent vectors to $\alpha$-curve and $\beta$-curve at point $P$ are respectively as follows:  
\begin{align}\label{ab}
\overrightarrow{a}=\dfrac{\partial \overrightarrow{r}}{\partial \alpha}, \ \ \ \ \ \overrightarrow{b}=\dfrac{\partial\overrightarrow{ r}}{\partial \beta}.
\end{align}
The total differential of $\overrightarrow{r}$ is 
\begin{align}\label{dr}
d \overrightarrow{r}=\overrightarrow{a}d\alpha+\overrightarrow{b}d\beta.
\end{align}
The first fundamental form (e.g., \cite{P}) is defined as the dot product of Eq. \eqref{dr} with itself:
\begin{align}\label{fff} \nonumber
(ds)^2=d\overrightarrow{r}\cdot d\overrightarrow{r}&= \big(\overrightarrow{a}d\alpha+\overrightarrow{b}d\beta\big)\cdot \big(\overrightarrow{a}d\alpha+\overrightarrow{b}d\beta\big) \\ &=  E(d\alpha)^2+2Fd\alpha d\beta+G(d\beta)^2,
\end{align}
where $E=\overrightarrow{a}\cdot \overrightarrow{a}$, $F=\overrightarrow{a}\cdot \overrightarrow{b}$ and $G=\overrightarrow{b}\cdot\overrightarrow{b}$ are known as the Gaussian fundamental quantities.

\begin{itemize}
\item
From Eq. \eqref{fff}, the distance between two arbitrary points $P_1$ and $P_2$ on the surface can be calculated: $$s=\int_{P_1}^{P_2}\sqrt{E(d\alpha)^2+2Fd\alpha d\beta+G(d\beta)^2}=\int_{P_1}^{P_2}\sqrt{E+2F\Big(\dfrac{d\beta}{d\alpha}\Big) +G\Big(\dfrac{d\beta}{d\alpha}\Big)^2} d\alpha .$$

\item
The angle between $\overrightarrow{a}$ and $\overrightarrow{b}$ is simply given by 
\begin{align}\label{angle}
\cos \theta=\dfrac{\overrightarrow{a}\cdot \overrightarrow{b}}{|\overrightarrow{a}|.|\overrightarrow{b}|}=\dfrac{F}{\sqrt{EG}}.
\end{align}

\item
Incremental area is the magnitude of the cross product of $\overrightarrow{a}d\alpha$ and $\overrightarrow{b}d\beta$, i.e., 
\begin{align}\label{area1}\nonumber
dA=|\overrightarrow{a}d\alpha \times \overrightarrow{b}d\beta|&=|\overrightarrow{a}d\alpha|.|\overrightarrow{b}d\beta|\sin \theta \\ \nonumber &= |\overrightarrow{a}|.|\overrightarrow{b}|\sin \theta d\alpha d\beta \\ \nonumber&=
\sqrt{E}\sqrt{G} \sqrt{1-\cos^2 \theta}d\alpha d\beta \\ \nonumber &= \sqrt{EG}\sqrt{\dfrac{EG-F^2}{EG}}d\alpha d\beta  \ \ \ \ \ \ \ \ \ \ \ \ \ {\rm from \ Eq. \ } \eqref{angle} \\  &=\sqrt{EG-F^2}d\alpha d\beta.
\end{align}
\end{itemize}

Since we are dealing with latitudes and longitudes on a spherical or spheroidal model of the Earth, the vectors $\overrightarrow{a}$ and $\overrightarrow{b}$ are orthogonal (meridians are normal to equator parallels). Also, in maps, we are dealing with the polar and Cartesian coordinate systems in which their axes are perpendicular. Thus, from Eq. \eqref{angle}, because $\cos 90^\circ =0$, one obtains $F=0$.  

Therefore, the first fundamental form \eqref{fff} in map projection will be deduced to the following form:
\begin{align}\label{fff1} 
(ds)^2=E(d\alpha)^2+G(d\beta)^2.
\end{align}

\begin{figure}[!htp] 
\centering
\includegraphics[height=8cm, width=10cm]{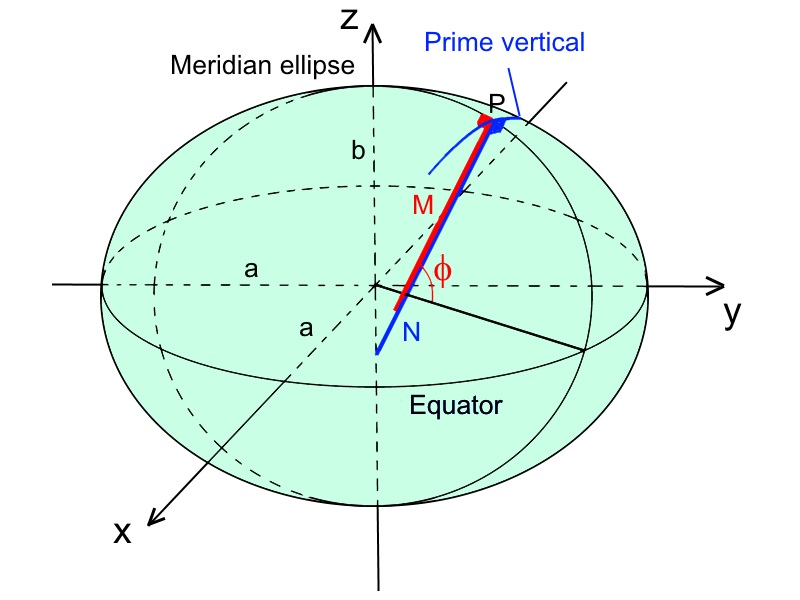}
\caption{  Geometry for the spheroidal model of the Earth, $\dfrac{x^2}{a^2}+\dfrac{y^2}{a^2}+\dfrac{z^2}{b^2}=1$. }
\label{ellips}
\end{figure}

\begin{example}{\label{ex1}
The first fundamental form for a planar surface

1. in the Cartesian coordinate system (a cylindrical surface) is $(ds)^2=(dx)^2+(dy)^2,$ where $E=G=1,$  

2. in the polar coordinate system (a conical surface) is $(ds)^2=(dr)^2+r^2(d\theta)^2,$ where $E=1$ and $G=r^2,$

3. in the spherical model of the Earth, Eq. \eqref{spherical}, is $(ds)^2=R^2(d\phi)^2+R^2\cos^2 \phi(d\lambda)^2,$ where $E=R^2$ and $G=R^2 \cos^2 \phi$, and

4. in the spheroidal model of the Earth, Eq. \eqref{spheroidal}, is $(ds)^2=M^2(d\phi)^2+N^2\cos^2 \phi(d\lambda)^2,$ where $E=M^2$ and $G=N^2 \cos^2 \phi$ in which $M$ is the radius of curvature in meridian and $N$ is the radius of curvature in prime vertical which are both functions of $\phi$: 
$$
M=\dfrac{a(1-e_{ab}^2)}{(1-e_{ab}^2 \sin^2\phi)^{1.5} }, \ \ \ \ \ \ \ \ N=\dfrac{a}{(1-e_{ab}^2\sin^2 \phi)^{0.5}}, \ \ \ \ \ e_{ab}^2=\dfrac{a^2-b^2}{a^2}.
$$
See Fig. \ref{ellips}, and \cite{P, VK} for the derivations of $M$ and $N$. }
\end{example}

Now suppose that $\phi$ and $\lambda$ are the parameters of the model of the Earth with the fundamental quantities $e$, $f$ and $g$. 

Consider a two-dimensional projection with parametric curves defined by the parameters $u$ and $v$. For instance, for the polar or conical coordinates, we have $u=r$ and $v=\theta$. Let $E',$  $ F'$ and $G'$ be its fundamental quantities.

Also, assume that on the plotting surface a second set of parameters, $x$ and $y$, with the fundamental quantities $E$, $F$ and $G$.

The relationship between the two sets of parameters on the plane is given by 
\begin{align}\label{eq1}
x=x(u,v), \ \ \ \ \ \ \ \ \ y=y(u,v).
\end{align} 
As an example, $x=r\cos\theta$ and $y=r\sin\theta$ for the polar and Cartesian coordinates. 

The relationship between the parametric curves $\phi$, $\lambda$, $u$ and $v$ is 
\begin{align}\label{eq2}
u=u(\phi, \lambda), \ \ \ \ \ \ \ \ \  v=v(\phi, \lambda).
\end{align}

Eq. \eqref{eq2} must be unique and reversible, i.e., a point on the Earth must represent only one point on the map and vice versa.
From Eqs. \eqref{eq1} and \eqref{eq2}, we have 
\begin{align}\label{xyuv}
x=x\big(u(\phi, \lambda), v(\phi, \lambda)\big), \ \ \ \ \ \ \ \ \ y=y\big(u(\phi, \lambda), v(\phi, \lambda)\big).
\end{align}

From the definition of the Gaussian first fundamental quantities, we have 

\begin{align}\label{efg}
E&= \overrightarrow{a}\cdot \overrightarrow{a}=\Big(\frac{\partial x}{\partial \phi}, \  \frac{\partial y}{\partial \phi}\Big)\cdot\Big(\frac{\partial x}{\partial \phi}, \ \frac{\partial y}{\partial \phi}\Big)=\Big(\frac{\partial x}{\partial \phi}\Big)^2+\Big(\frac{\partial y}{\partial \phi}\Big)^2, \\ \nonumber
F&=\overrightarrow{a}\cdot \overrightarrow{b}=\Big(\frac{\partial x}{\partial \phi}, \  \frac{\partial y}{\partial \phi}\Big)\cdot\Big(\frac{\partial x}{\partial \lambda}, \ \frac{\partial y}{\partial \lambda}\Big)=\frac{\partial x}{\partial \phi}\frac{\partial x}{\partial \lambda}+\frac{\partial y}{\partial \phi}\frac{\partial y}{\partial \lambda}, \\ \nonumber
G&=\overrightarrow{b}\cdot\overrightarrow{b}=\Big(\frac{\partial x}{\partial \lambda}, \  \frac{\partial y}{\partial \lambda}\Big)\cdot\Big(\frac{\partial x}{\partial \lambda}, \ \frac{\partial y}{\partial \lambda}\Big)=\Big(\frac{\partial x}{\partial \lambda}\Big)^2+\Big(\frac{\partial y}{\partial \lambda}\Big)^2.
\end{align}
Note that in here $\alpha$ and $\beta$ in \eqref{ab} are replaced by $\phi$ and $\lambda$, respectively. Similarly,  we have 
\begin{align}\label{efg'}
E'&= \overrightarrow{a}\cdot \overrightarrow{a}=\Big(\frac{\partial x}{\partial u}, \  \frac{\partial y}{\partial u}\Big)\cdot\Big(\frac{\partial x}{\partial u}, \ \frac{\partial y}{\partial u}\Big)=\Big(\frac{\partial x}{\partial u}\Big)^2+\Big(\frac{\partial y}{\partial u}\Big)^2, \\ \nonumber
F'&=\overrightarrow{a}\cdot \overrightarrow{b}=\Big(\frac{\partial x}{\partial u}, \  \frac{\partial y}{\partial u}\Big)\cdot\Big(\frac{\partial x}{\partial v}, \ \frac{\partial y}{\partial v}\Big)=\frac{\partial x}{\partial u}\frac{\partial x}{\partial v}+\frac{\partial y}{\partial u}\frac{\partial y}{\partial v}, \\ \nonumber
G'&=\overrightarrow{b}\cdot\overrightarrow{b}=\Big(\frac{\partial x}{\partial v}, \  \frac{\partial y}{\partial v}\Big)\cdot\Big(\frac{\partial x}{\partial v}, \ \frac{\partial y}{\partial v}\Big)=\Big(\frac{\partial x}{\partial v}\Big)^2+\Big(\frac{\partial y}{\partial v}\Big)^2.
\end{align}
As we mentioned earlier, since we are dealing with orthogonal curves, $f=F=F'=0$. Using this fact and Eqs. \eqref{xyuv}, \eqref{efg} and \eqref{efg'}, the following relation can be derived (see Section X Chapter 2 in \cite{P}):
\begin{align}\label{eg}
E=\Big(\frac{\partial u}{\partial \phi}\Big)^2 E'+\Big(\frac{\partial v}{\partial \phi}\Big)^2 G',\ \ \ \ \ \ \ \ \ \ 
G=\Big(\frac{\partial u}{\partial \lambda}\Big)^2 E'+\Big(\frac{\partial v}{\partial \lambda}\Big)^2 G'.
\end{align}
From Eq. \eqref{area1}, a mapping from the Earth to the plotting surface requires that 
\begin{align}\label{egeg}
eg=EG.
\end{align}
From Eqs. \eqref{efg}, \eqref{efg'}, \eqref{eg} and using $F=F'=0$, one obtains 
\begin{align}\label{Jacobian}
EG=J^2\cdot E'G', \ \ \ J=\begin{vmatrix} \dfrac{\partial u}{\partial \phi} & \dfrac{\partial u}{\partial \lambda} \\ \\ \dfrac{\partial v}{\partial \phi} & \dfrac{\partial v}{\partial \lambda}\end{vmatrix},
\end{align}
where $J$ is the Jacobian determinant of the transformation from the coordinate set $\phi$ and $\lambda$ to the coordinate set $u$ and $v$.

By a theorem of differential geometry (see \cite{P}), a mapping for the orthogonal curves  is conformal  if and only if 
\begin{align}\label{conformal}  
\frac{E}{e}=\frac{G}{g}.
\end{align}

\section{Projection from an ellipsoid to a sphere}\label{section4}
In this section, we describe how much the latitudes and longitudes of a spheroidal model of the Earth will be effected once they are transformed to a spherical model, i.e., how much distortion in shape and size happens when one projects a spheroidal model of the Earth to a spherical model, \cite{DA, P, RA, T}.
We distinguish two cases, equal-area transformation and conformal transformation. 

{\bf Case 1.} A spherical model of the Earth  that has the same surface area as that of the reference ellipsoid is called the {\it authalic sphere}. This sphere may be used as an intermediate step in the transformation from the ellipsoid to the mapping surface.  

Let $R_A$, $\phi_A$ and $\lambda_A$ be the authalic radius, latitude and longitude, respectively. Also, let $\phi$ and $\lambda$ be the geodetic latitude and longitude, respectively. 
From Example \ref{ex1}, we have $e=M^2$, $g=N^2\cos^2\phi$, $E'=R_A^2$ and $G'=R_A^2\cos^2\phi$. By Eqs. \eqref{egeg} and \eqref{Jacobian},
\begin{align}\label{authalic}
M^2N^2\cos^2\phi=R_A^4\cos^2\phi_A \begin{vmatrix} \dfrac{\partial \phi_A}{\partial \phi} & \dfrac{\partial \phi_A}{\partial \lambda} \\ \\ \dfrac{\partial \lambda_A}{\partial \phi} & \dfrac{\partial \lambda_A}{\partial \lambda}\end{vmatrix}^2.
\end{align}
In the transformation from the ellipsoid to the authalic sphere, longitude is invariant, i.e., $\lambda=\lambda_A$. Moreover, $\phi_A$ is independent of $\lambda_A$ and so $\lambda$. Thus Eq. \eqref{authalic} reduces to 
\begin{align}\label{authalic1}
M^2N^2\cos^2\phi=R_A^4\cos^2\phi_A \begin{vmatrix} \dfrac{\partial \phi_A}{\partial \phi} & 0\\ \\ 0 & 1\end{vmatrix}^2.
\end{align}
Substitute the values of $M$ and $N$ (given in Example \ref{ex1}) into Eq. \eqref{authalic1} to obtain 
\begin{align}\label{auth}
\dfrac{a^2(1-e_{ab}^2)}{(1-e_{ab}^2\sin^2\phi)^2}\cos\phi d\phi=R_A^2\cos\phi_Ad\phi_A.
\end{align} 
Integrating the left hand side of Eq. \eqref{auth} from $0$ to $\phi$ (using binary expansion), and the right hand side from $0$ to $\phi_A$, one obtains
\begin{align}\label{rasin}
R_A^2\sin \phi_A=a^2(1-e_{ab}^2)\Big(\sin\phi+\frac{2}{3}e_{ab}^2\sin^3\phi+\frac{3}{5}e_{ab}^4\sin^5\phi+\frac{4}{7}e_{ab}^6\sin^7\phi+\cdots\Big).
\end{align} 
Assuming $\phi_A=\pi/2$ when $\phi=\pi/2$, Eq. \eqref{rasin} gives: 
\begin{align}\label{ra}
R_A=a^2(1-e_{ab}^2)\Big(1+\frac{2}{3}e_{ab}^2+\frac{3}{5}e_{ab}^4+\frac{4}{7}e_{ab}^6+\cdots\Big).
\end{align}
Substituting Eq. \eqref{ra} into Eq. \eqref{rasin}, one obtains 
\begin{align}\label{sina}
\sin \phi_A=\sin\phi\bigg(\dfrac{1+\frac{2}{3}e_{ab}^2\sin^2\phi+\frac{3}{5}e_{ab}^4\sin^4\phi+\frac{4}{7}e_{ab}^6\sin^6\phi+\cdots}{1+\frac{2}{3}e_{ab}^2+\frac{3}{5}e_{ab}^4+\frac{4}{7}e_{ab}^6+\cdots}\bigg).
\end{align}
Since the eccentricity $e_{ab}$ is a small number, the above series are convergent. The relation between authalic and geodetic latitudes is equal at latitudes $0^\circ$ and $90^\circ$, and the difference between them at other latitudes is about $0^\circ.1$ for the WGS-84 spheroid (see \cite{P} for the definitions of the WGS-84 and WGS-72 spheroids).
\begin{example}{  

1. For the WGS-72 spheroid with $a\approx 6,378,135$  m and $e_{ab}\approx 0.081818$, the radius of the authalic sphere is $$R_A\approx a\sqrt{(1-e^2_{ab})\Big(1+\dfrac{2}{3}e_{ab}^2+\frac{3}{5}e_{ab}^4\Big)}\approx 6,371,004 \ {\rm m}.$$

2. For the I.U.G.G spheroid with $f=(a-b)/a\approx 1/298.275$, we have $e_{ab}=2f-f^2\approx 0.0066944$, and from Eq. \eqref{sina}, for geodetic latitude $\phi=45^\circ$, we have $\sin\phi_{A}\approx 0.70552$ which gives $\phi_A\approx 44^\circ.8713.$}

\end{example}

{\bf Case 2.} A conformal sphere is an sphere defined for conformal transformation from an ellipsoid, and similar to the authalic sphere may be used as an intermediate step in the transformation from the reference ellipsoid to a mapping surface.    

Let $R_c$, $\phi_c$ and $\lambda_c$ be the conformal radius, latitude and longitude for the conformal sphere, respectively. 
Let $e$ and $g$ be the same fundamental quantities as Case 1, and $E'=R_c^2$ and $G'=R_c^2\cos^2\phi_c$. Also, let $\phi_c=\phi_c(\phi)$ and $\lambda_c=\lambda$. 
Thus, from Eq. \eqref{eg}, 
\begin{align}\label{eg1}
E=\Big(\frac{\partial \phi_c}{\partial \phi}\Big)^2 E'+\Big(\frac{\partial \lambda_c}{\partial \phi}\Big)^2 G'=\Big(\frac{\partial \phi_c}{\partial \phi}\Big)^2 E',\ \ \ \ \ \ \ \ \ \ 
G=\Big(\frac{\partial \phi_c}{\partial \lambda}\Big)^2 E'+\Big(\frac{\partial \lambda_c}{\partial \lambda}\Big)^2 G'=G'.
\end{align}
Combining Eqs. \eqref{conformal} and \eqref{eg1}, one obtains $$\dfrac{\Big(\dfrac{\partial \phi_c}{\partial \phi}\Big)^2 R_c^2}{M^2}=\dfrac{R_c^2\cos^2\phi_c}{N^2\cos^2\phi},$$
that after integrating and simplifying with the condition $\phi_c=0$ for $\phi=0$, it gives 
\begin{align}\label{phic}
\tan\Big(\dfrac{\phi_c}{2}+\dfrac{\pi}{4}\Big)=\tan\Big(\dfrac{\phi}{2}+\dfrac{\pi}{4}\Big)\bigg(\dfrac{1-e_{ab}\sin\phi}{1+e_{ab}\sin\phi}\bigg)^{\dfrac{e_{ab}}{2}}.
\end{align}
One can calculate $\phi_c$ from Eq. \eqref{phic} which is a function of geodetic latitude $\phi$. Also, it can be shown that $R_c=\sqrt{MN}$ for a given latitude $\phi$ which in this case  $\phi=\pi/2$. We refer to Chapter 5 Section 3 in \cite{P} for the derivation.

\section{Albers and Lambert, one standard parallel}\label{section5}
In this section, we describe the Albers one standard parallel (equal-area conic projection) and Lambert one standard parallel (conformal conic projection) at latitude $\phi_0$ which give good maps around that latitude (cf., \cite{DFK, P, RA, T}).

We start with some geometric properties in a cone tangent to a spherical model of the Earth at latitude $\phi_0$.
\begin{figure}[!htp] 
\centering
\includegraphics[height=8cm, width=11cm]{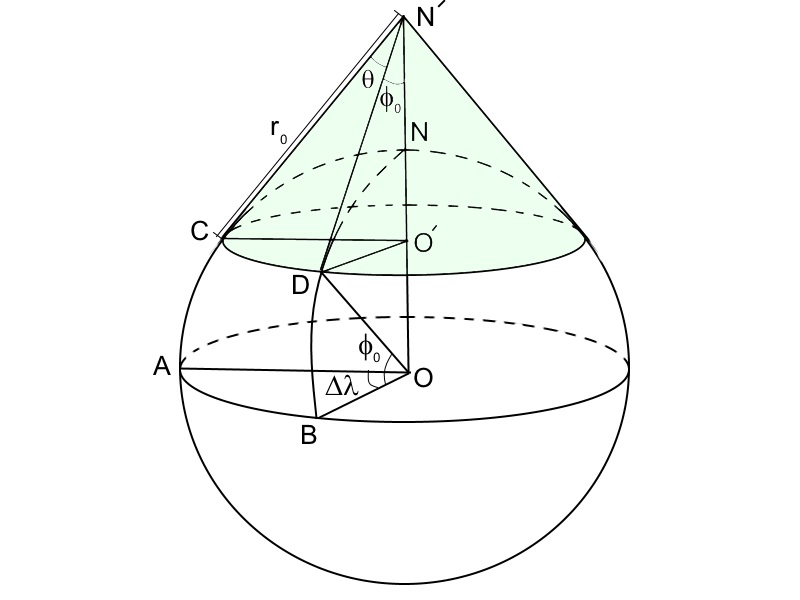}
\caption{  Geometry for angular convergence of the meridians}
\label{cone}
\end{figure}
In Fig. \ref{cone}, $ACN$ and $BDN$ are two meridians separated by  a longitude difference of $\Delta\lambda$, and $CD$ is an arc of the circle parallel to the equator. We have $CD=DO'\Delta\lambda$ and $DN'\sin\phi_0=DO'$ and approximately $\theta\cdot DN'=CD$. \\ Therefore, the first polar coordinate, $\theta$, is a linear function of $\lambda$, i.e., 
\begin{align}\label{theta}
\theta=\Delta\lambda\sin\phi_0.
\end{align}
The second polar coordinate, $r$, is a function of  $\phi$, i.e., 
\begin{align}\label{pcr}
r=r(\phi).
\end{align}
The constant of the cone, denoted $\varrho$, is defined from the relation between lengths on the developed cone on the Earth. Let the total angle on the cone, $\theta_T$, corresponding to $2\pi$ on the Earth be $\theta_T=d/r_0$, where $d=2\pi R\cos\phi_0$ is the circumference of the parallel circle to the equator at latitude $\phi_0$, and $r_0=R\cot\phi_0$. Thus $\theta_T=2\pi\sin\phi_0$, and the constant of the cone is defined as $\varrho=\sin\phi_0$.

\begin{figure}[!htp]
\centering
\begin{minipage}{.5\textwidth}
  \centering
  \includegraphics[height=6cm, width=7cm]{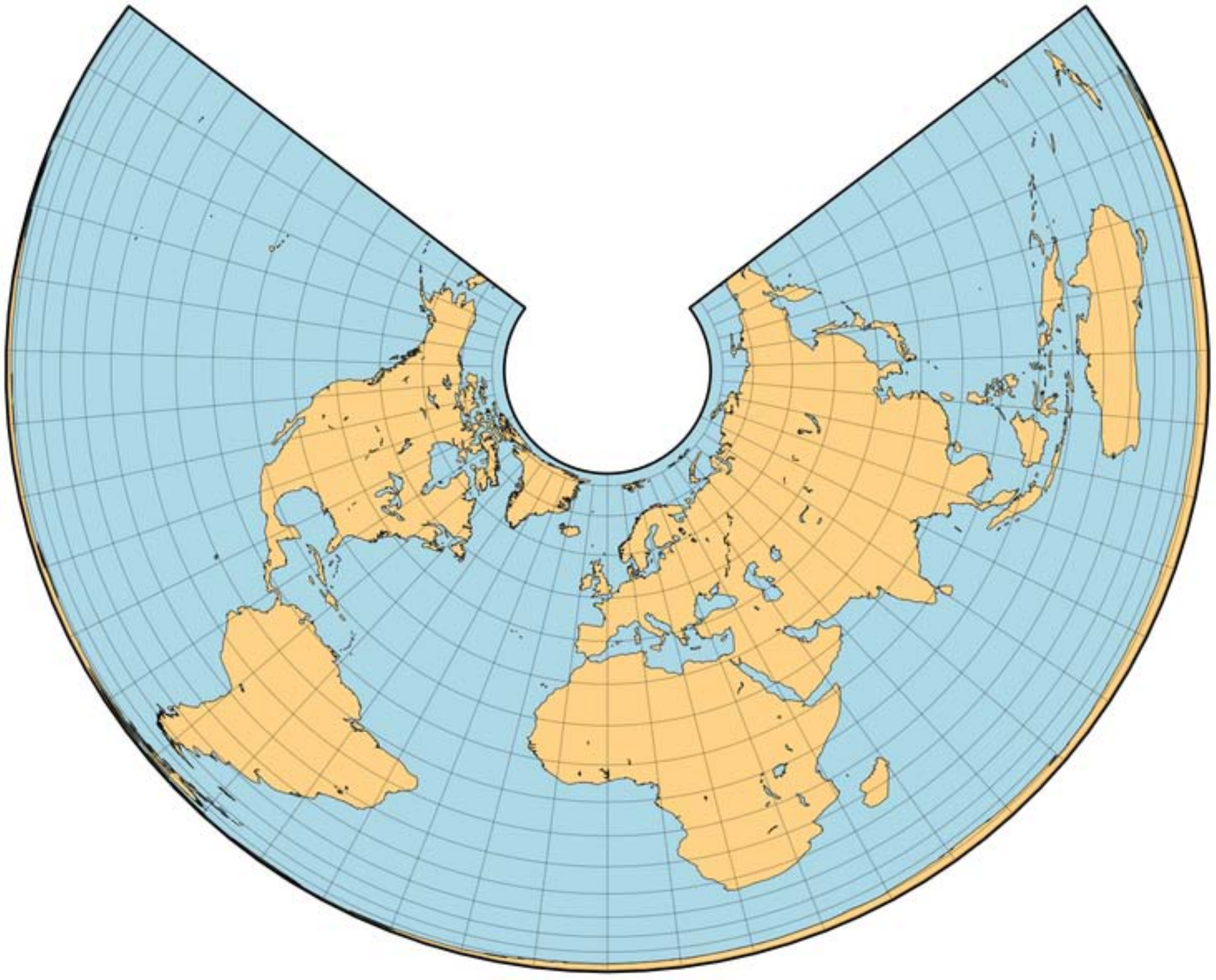}
  \caption{The Albers equal-area map with standard parallel  $45^{\circ}$N.}
  \label{al}
\end{minipage}%
\begin{minipage}{.5\textwidth}
  \centering
  \includegraphics[height=4cm, width=8.5cm]{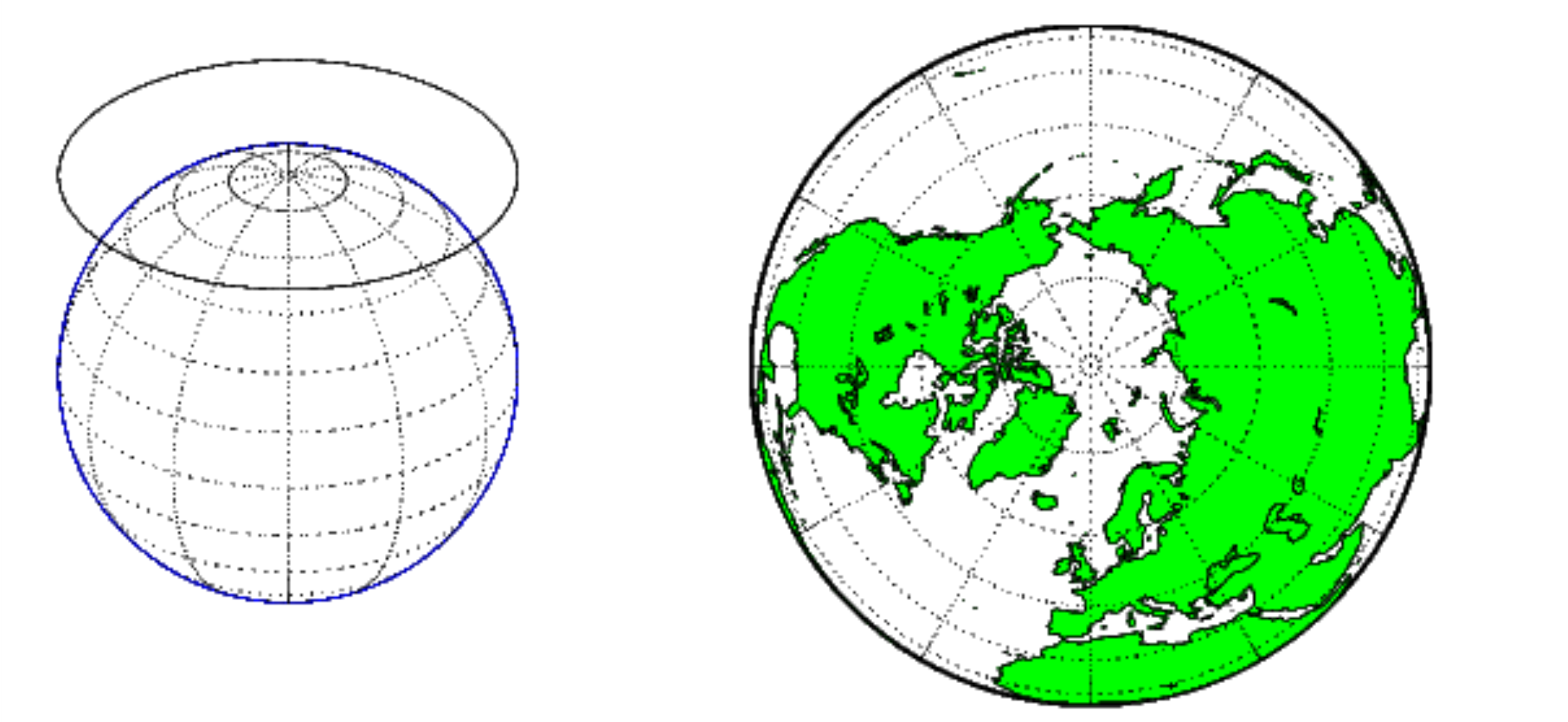}
  \caption{The Albers azimuthal map}
  \label{la}
\end{minipage}
\end{figure}

{\bf Case 1}. The Albers projection. Consider a spherical model of the Earth. From Example \ref{ex1}, we know that the first fundamental quantities for the sphere are 
$e=R^2$ and $g=R^2\cos^2\phi$ and for a cone (the polar coordinate system) are $E'=1$ and $G'=r^2$. Hence, from Eqs. \eqref{egeg} and \eqref{Jacobian}, 
\begin{align}\label{albers}
R^4\cos^2\phi=r^2 \begin{vmatrix} \dfrac{\partial r}{\partial \phi} & \dfrac{\partial r}{\partial \lambda}\\ \\ \dfrac{\partial \theta}{\partial \phi} & \dfrac{\partial \theta}{\partial \lambda}\end{vmatrix}^2.
\end{align} 
Using Eqs. \eqref{theta} and \eqref{pcr}, Eq. \eqref{albers} becomes  
\begin{align}\label{albers1}
R^4\cos^2\phi=r^2 \begin{vmatrix} \dfrac{\partial r}{\partial \phi} & 0\\ \\ 0 & \sin\phi_0\end{vmatrix}^2.
\end{align} 
Solving Eq. \eqref{albers1} by knowing the fact that an increase in $\phi$ corresponds to a decrease in $r$, one gets 
\begin{align}\label{ralbers}
r^2=\dfrac{-2R^2\sin\phi}{\sin\phi_0}+c.
\end{align}
Imposing the boundary condition $r_0=R\cot\phi_0$ into Eq. \eqref{ralbers}, $c=2R^2+R^2\cot^2\phi_0$, and so after some simplifications, Eq. \eqref{ralbers} becomes
\begin{align}\label{ralbers1}
r=\dfrac{R}{\sin\phi_0}\sqrt{1+\sin^2\phi_0-2\sin\phi\sin\phi_0}.
\end{align}
The Cartesian plotting equations for a conical projection are defined as follows:
\begin{align}\label{xys}
x=sr\sin\theta,\ \ \ \ \ \ \ \ y=s(r_0-r\cos\theta),
\end{align}
where $s$ is the scale factor, $\theta$ and $r$ are given respectively by Eqs. \eqref{theta} and \eqref{ralbers1}, and $r_0=R\cot\phi_0$. The origin of the projection has the coordinates $\lambda_0$ (the longitude of central meridian) and $\phi_0$. Fig. \ref{al} shows the Albers projection with one standard parallel.

If we let $\phi_0=90^\circ$, then Eqs. \eqref{theta} and \eqref{ralbers1} reduce to $$\theta=\Delta \lambda, \ \ \ \ \ \ \ \ \ \ r=R\sqrt{2(1-\sin\phi)},$$ that are the polar coordinates for the azimuthal equal-area projection, a special case of the Albers projection, see Fig. \ref{la}.

\begin{figure}[!htp]
\centering
  \centering
  \includegraphics[height=6cm, width=7.5cm]{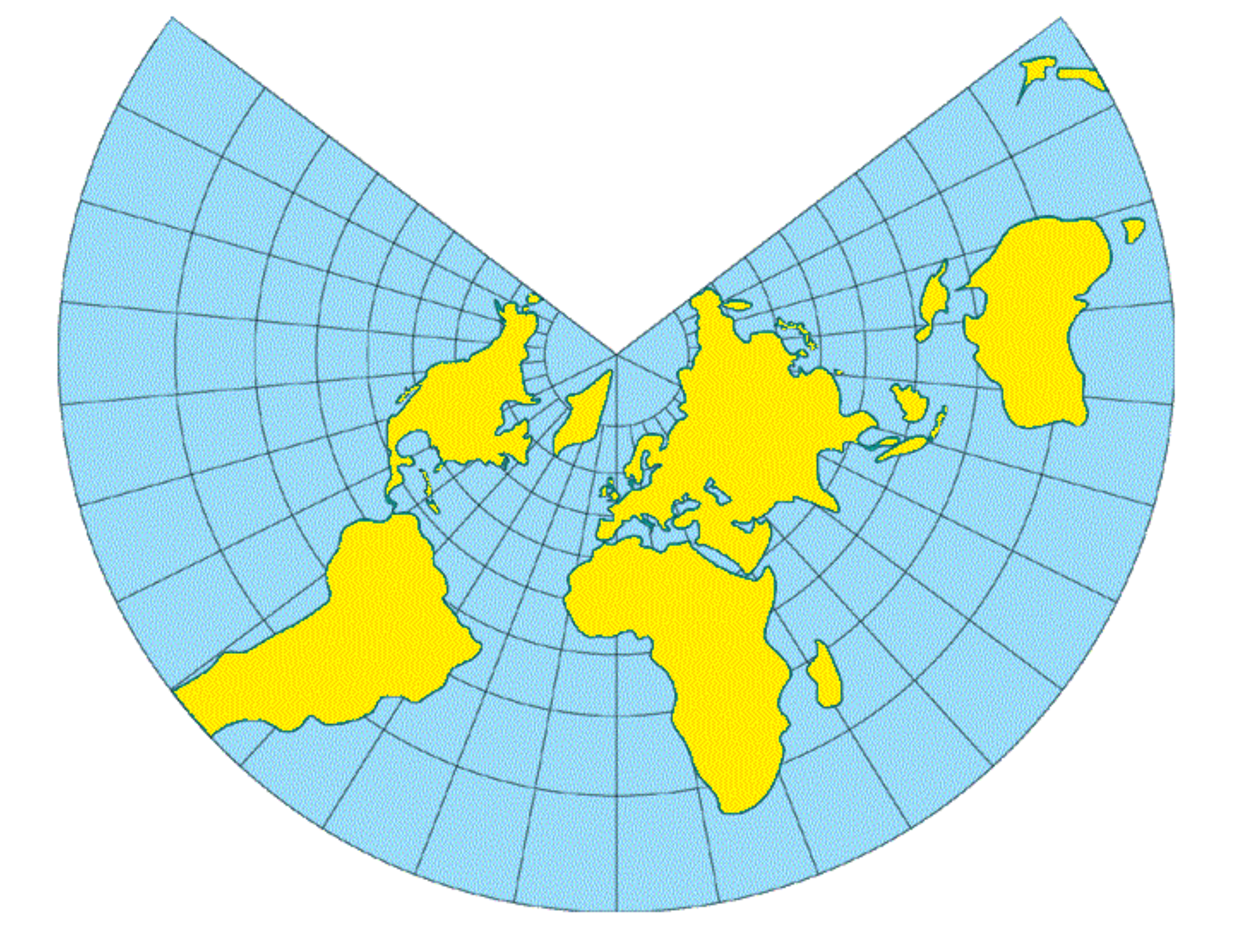}
  \caption{The Lambert conformal map}
  \label{Lamco}
\end{figure}

{\bf Case 2.} The Lambert projection. In this case, we consider a spheroidal model of the Earth. From Example \ref{ex1}, the fundamental quantities for this model are 
$e=M^2$ and $g=N^2\cos^2\phi$, and the fundamental quantities for a cone are $E'=1$ and $G'=r^2$. Again using Eqs.  \eqref{theta} and \eqref{pcr}, Eq. \eqref{eg} becomes 
\begin{align}\label{eg2}
E=\Big(\frac{\partial r}{\partial \phi}\Big)^2 E'+\Big(\frac{\partial \theta}{\partial \phi}\Big)^2 G'=\Big(\frac{\partial r}{\partial \phi}\Big)^2,\ \ \ \ \ \ \ \ \ \ 
G=\Big(\frac{\partial r}{\partial \lambda}\Big)^2 E'+\Big(\frac{\partial \theta}{\partial \lambda}\Big)^2 G'=\sin^2\phi_0 r^2.
\end{align}
Substituting these values in Eq. \eqref{conformal}, integrating, simplifying and noting that $r$ increases as $\phi$ decreases, one gets 
\begin{align}
r=r_0\left\{\dfrac{\tan\Big(\dfrac{\pi}{4}-\dfrac{\phi}{2}\Big)\bigg(\dfrac{1+e_{ab}\sin\phi}{1-e_{ab}\sin\phi}\bigg)^{e_{ab}/2}}{\tan\Big(\dfrac{\pi}{4}-\dfrac{\phi_0}{2}\Big)\bigg(\dfrac{1+e_{ab}\sin\phi_0}{1-e_{ab}\sin\phi_0}\bigg)^{e_{ab}/2}}\right\}^{\sin\phi_0},
\end{align} 
where $$r_0=N_{\phi_0}\cot\phi_0=\dfrac{a\cot\phi_0}{\big(1-e^2_{ab}\sin^2\phi_0\big)^{0.5}}.$$ The Cartesian equations are the same as Eq. \eqref{xys} with these new $r_0$ and $r$. Fig. \ref{Lamco} shows the Lambert projection with one standard parallel.


\section{Sinusoidal projection}\label{section6}
In this section, we only discuss about the sinusoidal equal-area projection that is a projection of the entire model of the Earth onto a single map, and it gives an adequate whole world coverage, \cite{DA, P}.

\begin{figure}[!htp] 
\centering
\includegraphics[height=6.2cm, width=11cm]{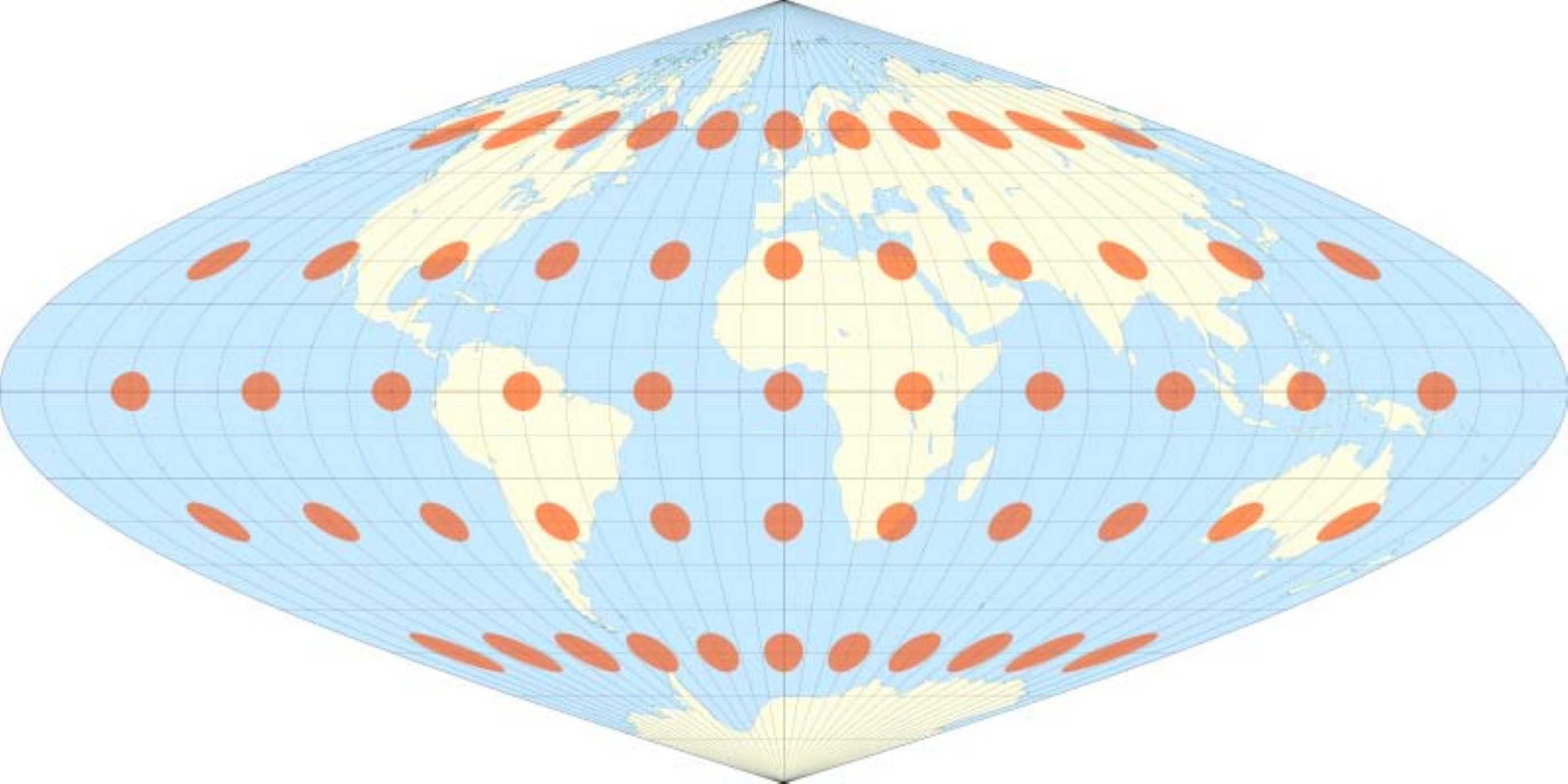}
\caption{The sinusoidal equal-area projection with Tissot's indicatrices that are changing  their shape (the ellipses with different eccentricities indicating angular distortion) toward the poles while having the same size. }
\label{sinusoidal1}
\end{figure}

Consider a spherical model of the Earth with the fundamental quantities $e=R^2$ and $g=R^2\cos^2\phi$. The first fundamental quantities on a planar mapping surface is $E'=G'=1$. Substituting these fundamental quantities into Eq. \eqref{Jacobian} (using Eq. \eqref{egeg}), one gets 
\begin{align*}
R^4\cos^2\phi=\begin{vmatrix} \dfrac{\partial x}{\partial \phi} & \dfrac{\partial x}{\partial \lambda}\\ \\ \dfrac{\partial y}{\partial \phi} & \dfrac{\partial y}{\partial \lambda}\end{vmatrix}^2,
\end{align*} 
which by imposing the conditions $y=R\phi$ and $x=x(\phi, \lambda)$ reduces to 
\begin{align}\label{yphi}
R^4\cos^2\phi=\begin{vmatrix} \dfrac{\partial x}{\partial \phi} & \dfrac{\partial x}{\partial \lambda}\\ \\ R & 0\end{vmatrix}^2=R^2\Big(\dfrac{\partial x}{\partial \lambda}\Big)^2.
\end{align} 
Taking the positive square root of Eq. \eqref{yphi} and using the fact that $\lambda$ and $\phi$ are independent, one obtains $dx=R\cos\phi d\lambda$, and so by integrating $x=\lambda R\cos\phi+c$. Using the boundary condition $x=0$ when $\lambda=\lambda_0$, one gets $c=-\lambda_0R\cos\phi$, and so the plotting equations for the sinusoidal projection become as follow ($\phi$ and $\lambda$ in radians):
\begin{align}\label{sinusoidal}
x=sR\Delta\lambda \cos\phi,\ \ \ \ \ y=sR\phi,
\end{align}
where $s$ is the scale factor. Fig. \ref{sinusoidal1} shows  a normalized plot for the sinusoidal projection. In this map, the meridians are sinusoidal curves except the central meridian which is a vertical line and they all meet each other in the poles. This is why this map is known as the sinusoidal map. The $x$ axis is also along the equator.

The inverse transformation from the Cartesian to geographic coordinates is simply calculated from Eq. \eqref{sinusoidal}: $$\phi=\dfrac{y}{sR}, \ \ \ \ \ \Delta\lambda=\dfrac{x}{sR\cos\phi}.$$

\section{Some conventional projections}\label{section7}
In this section, we give the  plotting equations for two conventional projections, the simple conic projection (one standard parallel) and the plate carree projection (cf., \cite{DA, P, S}). As we mentioned earlier, these projections neither preserve the shape nor do they preserve the size, and they are usually used  for simple portrayals of the world or regions with minimal geographic data such as index maps.

\begin{figure}[!htp] 
\centering
\includegraphics[height=5.5cm, width=7.5cm]{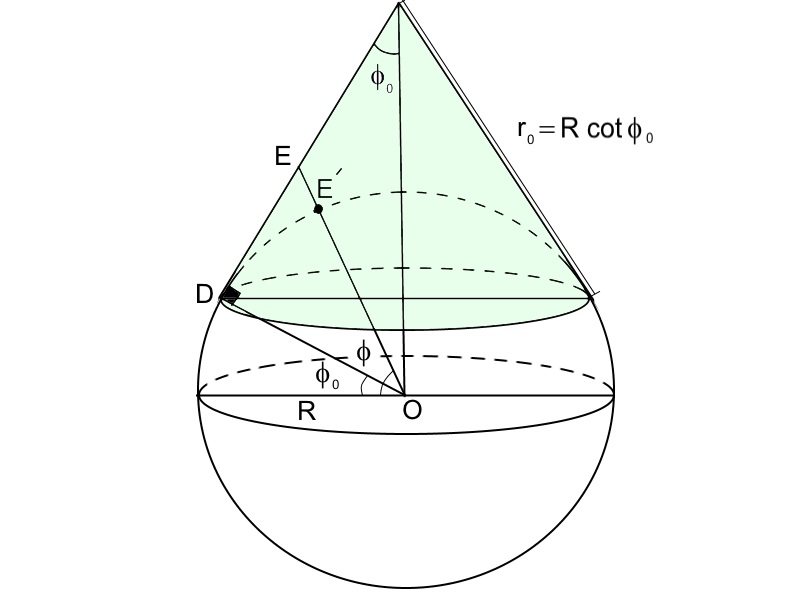}
\caption{  Geometry for the simple conic projection}
\label{conventional}
\end{figure}

{\bf 1.} The simple conic projection is a projection that the distances along every meridian are true scale. Suppose that the conic is tangent to the spherical model of the Earth at latitude $\phi_0$, see Fig. \ref{conventional}. In this figure, we have $r_0=R\cot\phi_0$. We want to have $DE=DE'$, but $DE'=R(\phi-\phi_0)$. Thus the polar coordinates for this projection are $$r=r_0-R(\phi-\phi_0), \ \ \ \ \ \ \theta=\Delta\lambda\sin\phi_0.$$ Replacing these values into Eq. \eqref{xys} gives its  Cartesian coordinates.

\begin{figure}[!htp] 
\centering
\includegraphics[height=6cm, width=11cm]{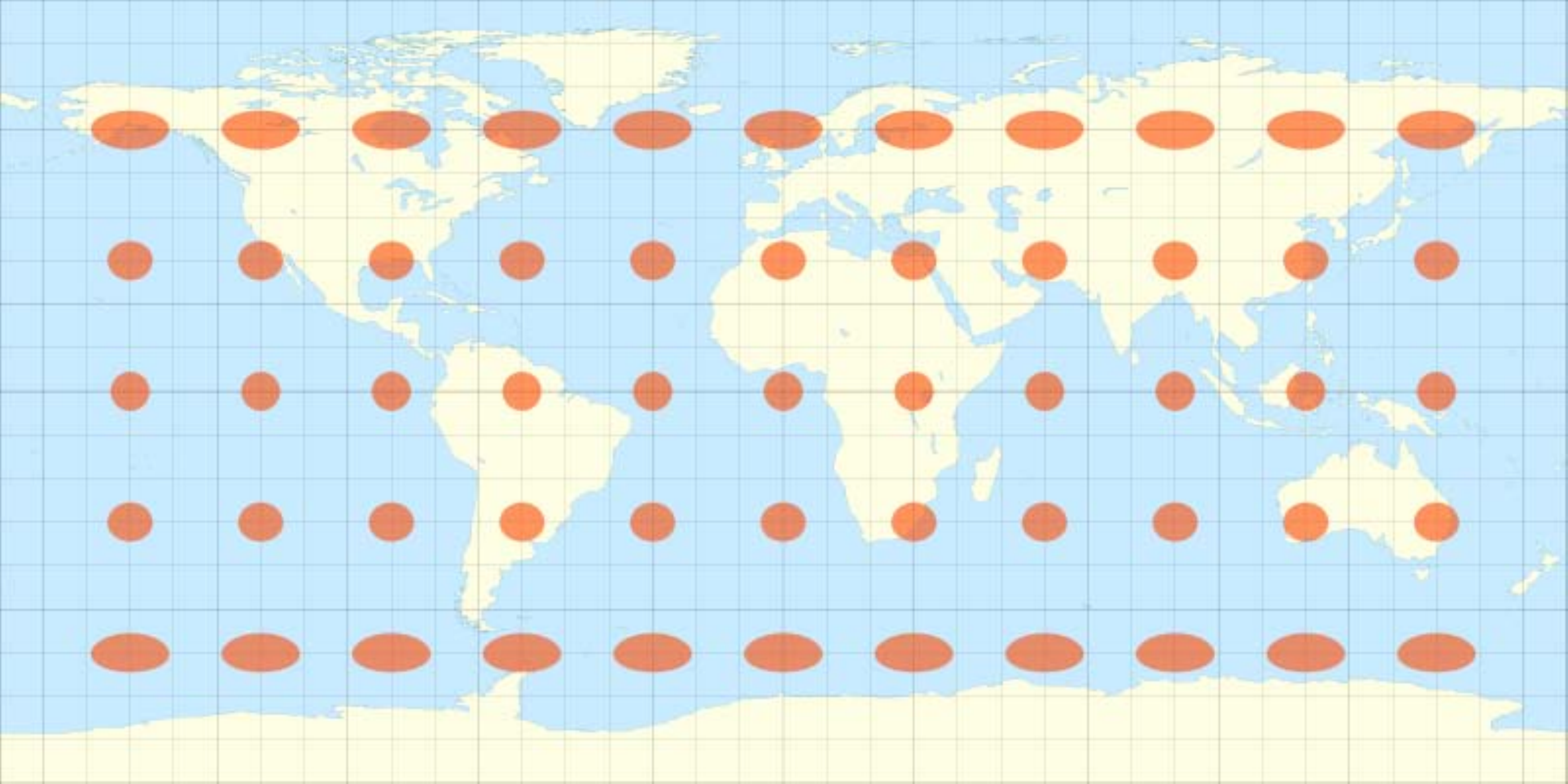}
\caption{ The plate carree map, $10^{\circ}$ graticule.}
\label{platec}
\end{figure}

{\bf 2.} The plate carree, the equirectangular projection, is a conventional cylindrical projection that divides the meridians equally the same way as on the sphere. Also, it divides the equator and its parallels equally. The plate carree plotting equations are very simple: $$x=sR\Delta\lambda, \ \ \ \ \ \ y=sR\phi,$$
where $\phi$ and $\lambda$ are in radians. Fig. \ref{platec} shows the plate carree map with Tissot's indicatrices which are changing their shape and size when moving toward the poles indicating that this map is neither equal-area nor conformal.
\section{Theory of distortion}\label{section8}
In this section, we discuss about three types of distortions from differential geometry approach: distortions in length, area and angle, and we present them in term of the Gaussian fundamental quantities (cf., \cite{P, RA, VK}).

{\bf 1.} The distortion in length is defined as the ratio of a length of a line on a map to the length of the true line on a model of the Earth. More precisely, 
\begin{align}\label{ldis}
K^2_L=\dfrac{(ds)^2_M}{(ds)^2_E}=\dfrac{E(d\phi)^2+G(d\lambda)^2}{e(d\phi)^2+g(d\lambda)^2}.
\end{align}
From Eq. \eqref{ldis}, the distortion along the meridians ($d\lambda=0$)  is $K_m=\sqrt{\dfrac{E}{e}}$, and along the lines parallel to the equator ($d\phi=0$) is $K_e=\sqrt{\dfrac{G}{g}}$.

{\bf 2.} The distortion in area is defined as the ratio of an area on a map to the true area on a model of the Earth. From Eq. \eqref{area1} ($f=F=0$), the area on the map is $A_M=\sqrt{EG}$, and the corresponding area on the model of the Earth is $A_E=\sqrt{eg}$. Thus, the distortion in area is 
\begin{align}\label{adis}
K_A=\dfrac{A_M}{A_E}=\sqrt{\dfrac{EG}{eg}}=K_mK_e.
\end{align}    
In equal-area map projections, from Eq. \eqref{egeg}, $K_A=K_mK_e=1$.

\begin{figure}[!htp] 
\centering
\includegraphics[height=8cm, width=10cm]{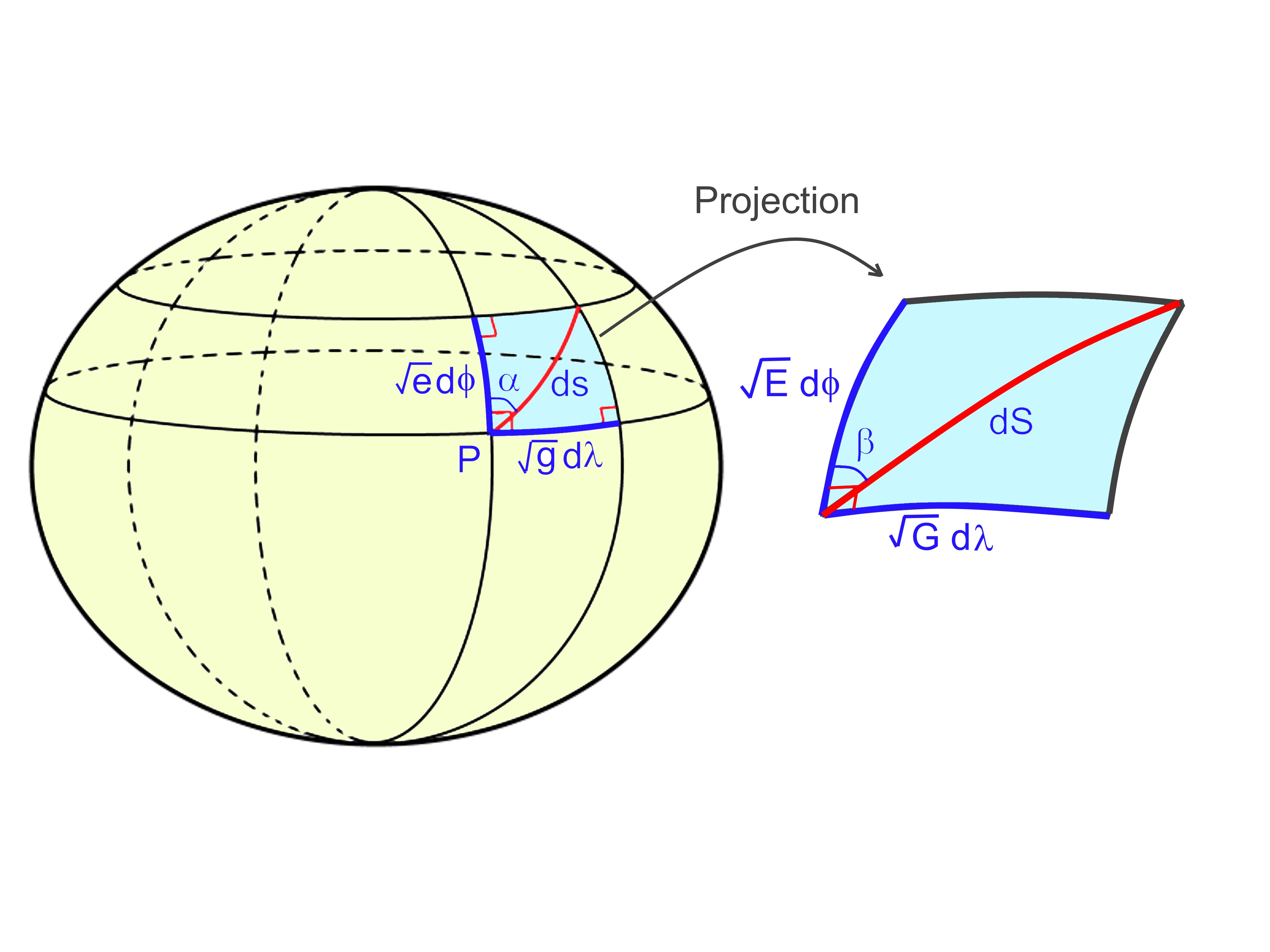}
\caption{  Geometry for differential parallelograms }
\label{distortion}
\end{figure}

{\bf 3.} The distortion in angle is defined as (in percentage):
\begin{align}\label{angledis}
K_{\alpha}=100\cdot\dfrac{\alpha-\beta}{\alpha},
\end{align}
where $\alpha$ is the angle on a model of the Earth (the azimuth), and $\beta$ is the projected angle on a map (the azimuth $\alpha$ on the map, cf., Fig. \ref{distortion}). 

In order to obtain $\beta$ as a function of the fundamental quantities and $\alpha$,  we first calculate $\sin(\beta\pm\alpha)$.
From Fig. \ref{distortion}, we have 
\begin{align*}
\sin(\beta\pm\alpha)&=\sin\beta\cos\alpha\pm \cos\beta\sin\alpha\\ &=\Big(\sqrt{G}\cdot\dfrac{d\lambda}{dS}\Big)\Big(\sqrt{e}\cdot\dfrac{d\phi}{ds}\Big)\pm\Big(\sqrt{E}\cdot\dfrac{d\phi}{dS}\Big)\Big(\sqrt{g}\cdot\dfrac{d\lambda}{ds}\Big)\\ &=(K_e\pm K_m)\dfrac{d\phi}{dS}\dfrac{d\lambda}{dS}\sqrt{eg}.
\end{align*}
Hence, $$\sin(\beta-\alpha)=\dfrac{K_e- K_m}{K_e+ K_m}\sin(\beta+\alpha).$$
Define $$f(\beta)=\sin(\beta-\alpha)-\dfrac{K_e- K_m}{K_e+ K_m}\sin(\beta+\alpha).$$ Now the goal is to find the roots of $f$. This can be done by 
Newton's iteration as follows:
\begin{align*}
\beta_{n+1}=\beta_n-\dfrac{f(\beta_n)}{f'(\beta_{n})},
\end{align*}
where $$f'(\beta_n)=\cos(\beta_n-\alpha)-\dfrac{K_e- K_m}{K_e+ K_m}\cos(\beta_n+\alpha).$$
The iteration is rapidly convergent by letting $\beta_0=\alpha$.
In conformal mapping, from Eq. \eqref{conformal}, $K_e=K_m$, and so the function $f$ will have a unique solution ($\beta=\alpha$). 

\begin{example} {In this example, we show the distortions in length in the Albers projection with one standard parallel. 
From Example \ref{ex1}, the first fundamental form for the map is 
\begin{align}\label{dsm}
(ds)^2_M=(dr)^2+r^2(d\theta)^2=E'(dr)^2+G'(d\theta)^2,
\end{align}
and the first fundamental form for the spherical model of the Earth is
\begin{align}\label{dse}
(ds)^2_E=R^2(d\phi)^2+R^2\cos^2\phi(d\lambda)^2=e(d\phi)^2+g(d\lambda)^2.
\end{align}
Taking the derivatives of Eqs. \eqref{theta} and \eqref{albers1}, one obtains $$d\theta=\sin\phi_0d\lambda, \ \ \ \ \ \ \ \ \ dr=\dfrac{-R\cos\phi \ d\phi}{\sqrt{1+\sin^2\phi_0-2\sin\phi\sin\phi_0}},$$ respectively.  Substitute the above equations into Eq. \eqref{dsm} to get 
\begin{align}\label{dsmm}
(ds)^2_M=\dfrac{R^2\cos^2\phi }{1+\sin^2\phi_0-2\sin\phi\sin\phi_0}(d\phi)^2+r^2\sin^2\phi_0(d\lambda)^2=E(d\phi)^2+G(d\lambda)^2.
\end{align}
Substituting \eqref{dse} and \eqref{dsmm} in Eq. \eqref{ldis} gives the total length distortion. Also, $$K_m=\sqrt{\dfrac{E}{e}}=\dfrac{\cos\phi }{\sqrt{1+\sin^2\phi_0-2\sin\phi\sin\phi_0}}, \ \ \ \ \ \ K_e=\sqrt{\dfrac{G}{g}}=\dfrac{\sqrt{1+\sin^2\phi_0-2\sin\phi\sin\phi_0}}{\cos\phi },$$ which are functions of $\phi$. Clearly, $K_mK_e=1.$
}\end{example}
\begin{example}{
In this example, we first use the first fundamental form to obtain the plotting equations for the Mercator projection, and then we show its length and area distortion. 
From Example \ref{ex1}, the first fundamental form for the cylindrical surface (the Cartesian coordinate system) is  
\begin{align}\label{cyxy}
(ds)_M^2=(dy)^2+(dx)^2.
\end{align}
Taking the derivative of Eq. \eqref{xymercator} and substituting in Eq. \eqref{cyxy}, one finds 
\begin{align}\label{cyxy1}
(ds)_M^2=\Big(\dfrac{dy}{d\phi}\Big)^2(d\phi)^2+s^2R^2(d\lambda)^2=E(d\phi)^2+G(d\lambda)^2,
\end{align}
where $s$ is the scale of the map along the equator, $E=\big(dy/d\phi\big)^2$ and $G=s^2R^2.$
The first fundamental quantities for the spherical model of the Earth are $e=R^2$ and $g=R^2\cos^2\phi$.
Substituting these fundamental quantities in Eq. \eqref{conformal} and simplifying, one obtains 
\begin{align}\label{dydphi}
dy=\dfrac{sRd\phi}{\cos\phi}.
\end{align}
It is easy to see that integrating the above differential equation and applying the boundary condition $y(0)=0$, Eq. \eqref{ymer} follows.  By Eq. \eqref{dydphi},
$E=\big(dy/d\phi\big)^2=s^2R^2/\cos^2\phi.$ Therefore, substituting Eqs. \eqref{dse} and \eqref{cyxy1} in Eq. \eqref{ldis}, the length distortion will be $$K_L=\dfrac{s}{\cos\phi}.$$
It can be seen that $K_L=K_m=K_e$, and so from Eq. \eqref{adis}, the distortion in area for the Mercator projection is $$K_A=K_mK_e=\dfrac{s^2}{\cos^2\phi}.$$
Hence, in the Mercator projection both length and area distortions are functions of $\phi$ not $\lambda$. 
}\end{example}

\section{Conclusion}

There are a number of map projections used for different purposes, and we discussed about three major classes of them, equal-area, conformal, and conventional. Users may also create their own map based on their projects by starting with a base map of known projection and scale.  

In this paper, in cylindrical projections, we assume that the cylinder is tangent to the equator. Making the cylinder tangent to  other closed curves  on the Earth results good maps in areas close to the tangency. This is also applied for conical and azimuthal projections.

In all projections from a 3-D surface to a 2-D surface, there are distortions in length, shape or size that some of them can be removed (not all) or minimized from the map based on some specific applications. We also noticed in Section \ref{section4} that projecting a spheroidal model of the Earth to a  spherical model of the Earth will also distort length, shape and angle.

Intelligent map users should have knowledge about the theory of distortion in order to compare and distinguish their maps with the true surface on the Earth that they are studying. 

{}

\end{document}